\documentclass[lettersize,journal]{IEEEtran}
\usepackage{cite}
\usepackage{amsmath, amssymb, amsfonts}
\usepackage{algorithmic}
\usepackage{graphicx}
\usepackage{textcomp}
\usepackage{xcolor}
\usepackage{graphicx}
\usepackage{epstopdf}
\usepackage{algorithm}
\usepackage{algorithmic}
\usepackage{bbm}
\usepackage{bm}
\usepackage{mathrsfs}
\usepackage{graphicx}
\usepackage{subcaption}
\usepackage{caption}  
\usepackage{xcolor}
\usepackage{xpatch}

\makeatletter
\def\changeBibColor#1{
	\in@{#1}{NP,Genetic}
    
	\ifin@\color{black}\else\normalcolor\fi
}
\xpatchcmd\@bibitem
{\item}
{\changeBibColor{#1}\item}
{}{}
\makeatother

\newtheorem{proposition}{\underline{Proposition}}
\begin{document}
\belowdisplayskip=1.5pt
\title{Handover-Aware Trajectory Optimization for Cellular-Connected UAV}

\author{\IEEEauthorblockN{Xiangming Du, Shuowen Zhang, \emph{Senior Member, IEEE}, and Francis C.-M. Lau, \emph{Fellow, IEEE}}\vspace{-9mm}\\
\thanks{This work was supported in part by the National Natural Science Foundation of China under Grant 62101474, in part by the General Research Fund from the Hong Kong Research Grants Council under Grant 15230022, and in part by the Young Collaborative Research Grant from the Hong Kong Research Grants Council under Grant PolyU C5002-23Y. The authors are with the Department of Electrical and Electronic Engineering, The Hong Kong Polytechnic University, Hong Kong SAR, China (e-mail: xiangming.du@connect.polyu.hk; shuowen.zhang@polyu.edu.hk; francis-cm.lau@polyu.edu.hk).}
}
\maketitle

\begin{abstract}
In this letter, we study a cellular-connected unmanned aerial vehicle (UAV) which aims to complete a mission of flying between two pre-determined locations while maintaining satisfactory communication quality with the ground base stations (GBSs). Due to the potentially long distance of the UAV's flight, frequent handovers may be incurred among different GBSs, which leads to various practical issues such as large delay and synchronization overhead. To address this problem, we investigate the trajectory optimization of the UAV to minimize the number of GBS handovers during the flight, subject to a communication quality constraint and a maximum mission completion time constraint. Although this problem is non-convex and difficult to solve, we derive useful structures of the optimal solution, based on which we propose an efficient algorithm based on graph theory and Lagrangian relaxation for finding a high-quality suboptimal solution in polynomial time. Numerical results validate the effectiveness of our proposed trajectory design.
\end{abstract}
\vspace{-2mm}
\vspace{-5mm}
\section{Introduction}
\vspace{-1mm}
\IEEEPARstart{I}{n} recent years, unmanned aerial vehicles (UAVs) have been deployed in military, civil, and many other applications due to their high-altitude operation capabilities. To ensure the safety of UAVs, a promising solution is \emph{cellular-enabled UAV communication} or \emph{cellular-connected UAV}, where UAVs will be served by the ground base stations (GBSs) in the cellular network as new aerial users \cite{ShuowenCellular}. Compared with traditional Wi-Fi based UAV-ground communication approaches, cellular-enabled UAV communication can extend the coverage range to beyond visual line-of-sight (LoS) by leveraging the high-speed backhaul links among the cellular GBSs \cite{ShuowenCellular}.

To maintain satisfactory communication quality with the GBSs, the UAV's trajectory needs to be carefully designed. Particularly, there exists a non-trivial trade-off between the communication quality and the mission completion performance (e.g., completion time), since the UAV should generally fly near the GBSs to enhance the communication link quality, which, on the other hand, may result in detoured paths in completing the mission. \cite{ShuowenCellular} studied the trajectory optimization for minimizing the mission completion time subject to a constant communication quality constraint throughout the flight, and proposed an efficient polynomial-time algorithm for finding an approximate solution with arbitrarily low performance gap with the optimal solution. \cite{ShuowenOutage} extended this work by allowing tolerable connection disruptions with the GBSs, while \cite{ShuowenRadio, 5GCellular} took the interference issue into consideration. On the other hand, \cite{GuoUAV} studied the communication-constrained trajectory design for the specific mission of information collection.

However, a critical practical issue overlooked in the existing literature lies in the potentially frequent \emph{handovers} among multiple GBSs that consecutively associate with the UAV. Due to the generally long distances of the UAV's flight, the UAV may have to frequently change the associated GBS, which leads to heavy signaling overhead and affects the service continuity \cite{Weipaper1}. In addition, the ping-pong effect caused by frequent handovers will downgrade the network security \cite{Handoverprotocol}. Experiments have shown that increased number of handovers leads to more frequent delay peaks \cite{Handovertest}. Although the handover issues (e.g., handover failure probability and ping-pong probability) have been analyzed and studied in e.g., \cite{Weipaper1, Handoverprobability}, how to minimize the number of handovers via UAV trajectory design is still an unaddressed open problem. Compared with other performance metrics such as the mission completion time, the number of handovers is a discrete-valued function that is difficult to be characterized explicitly with respect to the UAV's trajectory. Moreover, there exists non-trivial trade-offs among the handover performance, mission completion performance, and communication performance, which make handover-aware trajectory optimization more challenging.

In this letter, we consider a cellular-connected UAV which needs to fly between two locations. To ensure its safety, it needs to maintain a satisfactory communication quality constraint with a GBS at every time instant. We aim to optimize the UAV's trajectory to minimize the number of handovers during the flight, subject to the communication quality constraint and a maximum threshold on the mission completion time. This problem is non-convex and difficult to solve. By judiciously exploring the problem structure, we transform the problem into a more tractable form, based on which we propose a polynomial-time algorithm by applying graph theory and Lagrangian relaxation to find a high-quality suboptimal solution. It is shown via numerical results that our proposed design requires fewer handovers compared with handover-unaware trajectory designs.
\vspace{-4mm}
\section{System Model}\vspace{-1mm}

We consider a cellular-enabled UAV communication system with $M \geq 1$ \emph{heterogeneous} GBSs and a UAV. Both the UAV and each GBS are equipped with one single antenna.\footnote{It is worth noting that our results can be readily extended to the case with multiple antennas at the GBSs/UAV by re-quantifying the communication quality with multi-antenna gain/beamforming taken into account.} We assume that the UAV flies at a constant height of $H$ in meters (m), and its three-dimensional (3D) location at time instant $t$ is denoted as $(x(t),y(t),H),\ 0\leq t\leq T$. Specifically, $T$ represents the completion time of the UAV's mission in seconds (s), which should be no longer than a pre-defined maximum mission completion time threshold $T_{\max}$. The mission of the UAV is to fly from an initial point $U_0$ to a final point $U_{\mathrm{F}}$, for which the locations are denoted by $[x_0,y_0,H]^T$ and $[x_{\mathrm{F}},y_{\mathrm{F}},H]^T$, respectively. During the flight, the UAV is required to maintain satisfactory downlink communication quality with the GBSs at every time instant.

\begin{figure}[t]
	\centering
	\includegraphics[width=5cm]{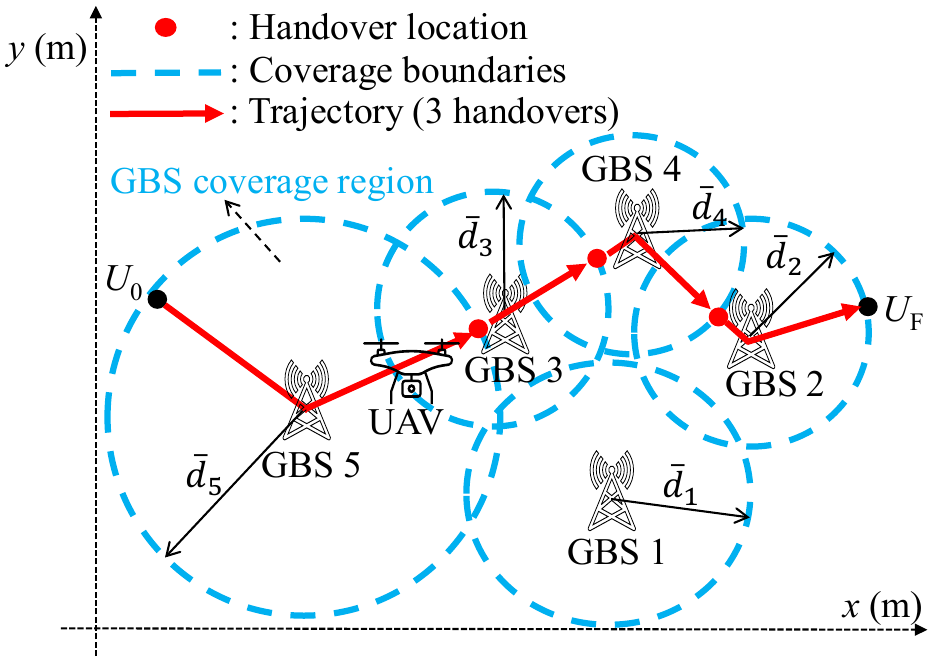}
	\vspace{-1mm}
	\caption{Illustration of handovers among different GBSs for a cellular-connected UAV.}\vspace{-8mm}
	\label{fig_1}
\end{figure}

We consider a \emph{heterogeneous} network where each GBS may have a different antenna height and a different transmit power for downlink UAV communication. For each $m$-th GBS, we let $H_m$ and $P_m$ denote the antenna height and transmit power for UAV communication, respectively, and $(a_m,b_m, H_m)$ denote its 3D location. For illustration, we further define $\bm{g}_m = [a_m,b_m]^T$ and $\bm{u}(t)=[x(t),y(t)]^T$ to represent the horizontal locations of each $m$-th GBS and the UAV at time instant $t$, respectively. In addition, we define $\bar{\bm{u}}_0=[x_0,y_0]^T$ and $\bar{\bm{u}}_{\mathrm{F}}=[x_{\mathrm{F}},y_{\mathrm{F}}]^T$, with $\bm{u}(0)=\bar{\bm{u}}_0$ and $\bm{u}(T)=\bar{\bm{u}}_{\mathrm{F}}$. We consider a maximum speed constraint of the UAV denoted by $V_{\max}$, which yields $\|\dot{\bm{u}}(t)\|\leq V_{\max}$. The link distance between the $m$-th GBS and the UAV at each time instant $t$ is given by $d_m(t)=\sqrt{(H-H_m)^2+\|\bm{u}(t)-\bm{g}_m\|^2}, \ m\in \mathcal{M}$, where $\mathcal{M}=\{1, \cdots\!, M\}$. For ease of understanding the fundamental trade-offs among handover, mission completion time, and communication quality, we consider the LoS channel model between GBSs and the UAV. Let $h_m(t)\in \mathbb{C}$ denote the complex baseband equivalent channel coefficient from each $m$-th GBS to the UAV at time instant $t$. The corresponding channel power gain is modeled as $|h_m(t)|^2=\frac{\beta_0}{d_{m}^{2}(t)}=\frac{\beta_0}{(H-H_m)^2+\|\bm{u}(t)-\bm{g}_m\|^2}, \ m\in \mathcal{M}$, where $\beta_0$ denotes the channel power gain at reference distance $d_0=1$ m. We assume that the UAV communicates with one GBS indexed by $I(t)\in \mathcal{M}$ at each time instant $t$, and a dedicated time-frequency resource block is allocated for UAV communication. The receive SNR at the UAV at time instant $t$ is thus given by\vspace{-3mm}
\begin{align}
\!\!\rho_{I(t)}(t)=\frac{P_{I(t)}\beta_0}{\sigma^2((H\!-\!H_{I(t)})^2\!+\!\| \bm{u}(t)\!-\!\bm{g}_{I(t)} \| ^2)},\ 0\!\leq \!t\!\leq \!T,
\end{align}
where $\sigma^2$ denotes the average noise power at the UAV receiver. We consider a communication quality requirement specified by a minimum receive SNR threshold denoted by $\bar{\rho}$. Namely, the UAV can satisfactorily communicate with GBS $I(t)$ at time instant $t$ if and only if $\rho_{I(t)}(t)\geq \bar{\rho}$ or equivalently $\|\bm{u}(t)-{\bm{g}_{I(t)}}\|\leq \bar{d}_{I(t)}\triangleq\sqrt{\frac{P_{I(t)}\beta_0}{\sigma^2\bar{\rho}}-(H-H_{I(t)})^2}$ holds, i.e., the horizontal location of the UAV lies in the disk-shaped (horizontal) \emph{coverage region} of GBS $I(t)$ centered at $\bm{g}_{I(t)}$ with radius $\bar{d}_{I(t)}$, as illustrated in Fig. \ref{fig_1}.

Note that at each time instant $t$, there may exist multiple GBSs that can satisfy the communication constraint when associated with the UAV. However, frequent change of the GBS-UAV association $I(t)$ over time leads to frequent GBS \emph{handovers}, and consequently causes increased delay and overhead (e.g., for synchronization). Motivated by this, we aim to judiciously design the UAV's trajectory and GBS-UAV associations $I(t)$'s to minimize the total number of handovers during the UAV's mission, while maintaining satisfactory communication quality and mission completion time.

To this end, we first mathematically characterize the number of handovers. We consider that a handover happens when the UAV reaches the boundary of the coverage region of the GBS currently associated with the UAV, and is about to enter the coverage region of another GBS, as illustrated in Fig. \ref{fig_1}. Specifically, define $\chi(t)$ to indicate whether the UAV has reached the boundary of the coverage region of the currently associated GBS:
\vspace{-1mm}\begin{equation}
\!\chi(t)\!=\!\!\sum_{t'\in \{\tau:\|\bm{u}(\tau)-\bm{g}_{I(\tau)}\|-\bar{d}_{I(\tau)}=0,\ \! \tau\in [0,T]\}}\!\!\!\!\delta(t\!-\!t'),\ 0\!\leq\! t\!\leq \!T,\!\!\!\!
\end{equation}
where $\delta(\cdot)$ denotes the Dirac-delta function. Moreover, let $\psi(t)$ indicate whether the UAV is ready to enter the coverage region of another GBS, which is given by
\vspace{-1mm}\begin{equation}
\!\!\!\!\psi(t)\!=\!\left\{
\begin{array}{l}
\!\!1,\ \text{if}\ \!\! \min\limits_{m\in \mathcal{M}\backslash \{I(t)\}}\!\|\bm{u}(t)\!-\!{\bm{g}_{m}}\|\!\!-\! \bar{d}_{m}\!\leq\! 0\\
\!\!0,\ \text{otherwise}
\end{array}\right.\!\!\!\!,\ \!\!0\!\leq\! t\!\leq\! T.
\vspace{-1mm}\end{equation}
Note that $\psi(t)=1$ indicates that the UAV is ready to be associated with another GBS indexed by $m\in \mathcal{M}\backslash \{I(t)\}$ that is different from the associated one, and $\psi(t)=0$ otherwise. Notice that $\chi(t)\psi(t)$ consists of multiple time-shifted Dirac-delta functions, each taking a non-zero value at a handover time instant. Therefore, the total number of handovers during the UAV's flight can be expressed as:\vspace{-2mm}
\begin{align}
N=\int_0^T\chi(t)\psi(t)dt.\label{handover function}
\end{align}

Note that the number of handovers in (\ref{handover function}) is critically dependent on the UAV's trajectory $\{\bm{u}(t),0\leq t\leq T\}$ and GBS-UAV associations $\{I(t),0\leq t\leq T\}$, which also critically affect the communication quality and mission completion time. In general, there exists a non-trivial trade-off among the handover, communication, and mission completion performances. For example, to maximize the receive SNR, the UAV should fly near the GBSs during its flight as much as possible, which, on the other hand, may result in a large number of handovers and long mission completion time. To resolve this trade-off, we will jointly optimize the UAV's trajectory and GBS-UAV associations for minimizing the number of handovers subject to communication quality and mission completion requirements.\vspace{-4mm}
\section{Problem Formulation}\vspace{-1mm}
We aim to jointly optimize the UAV's trajectory $\{\bm{u}(t),0\leq t\leq T\}$ and the GBS-UAV associations $\{I(t),0\leq t\leq T\}$ to minimize the number of handovers, subject to a communication quality constraint specified by a minimum receive SNR threshold $\bar{\rho}$ and a maximum mission completion time threshold $T_{\max}$. The optimization problem is formulated as\vspace{-3mm}
\begin{align} \mbox{(P1)}\ 
\mathop{\min_{\substack{T, \{\bm{u}(t),0\leq t\leq T\},\\ \{I(t),0\leq t\leq T\}}}} \ & N\\[-1mm]
\mathrm{s.t.}\quad\quad \ &\bm{u}(0) =\bar{\bm{u}}_0, \ \bm{u}(T) =\bar{\bm{u}}_{\mathrm{F}}\label{P1-C1}\\[-1mm]
&\|\bm{u}(t)\!-\!\bm{g}_{I(t)}\|\!-\!\bar{d}_{I(t)}\!\leq \!0,\ 0\!\leq \!t\!\leq\! T\!\label{P1-C2}\\[-1mm]
&I(t)\in \mathcal{M},\ 0\leq t\leq T\label{P1-C3}\\[-1mm]
&\|\dot{\bm{u}}(t)\|\leq V_{\max}, \ 0\leq t\leq T\label{P1-C4}\\[-1mm]
&T\leq T_{\max}.\label{P1-C5}
\end{align}

\vspace{-1mm}Note that the objective function defined in (\ref{handover function}) is in a complex form which is a non-decreasing step function of the mission completion time $T$. Moreover, both $\bm{u}(t)$ and $I(t)$ are continuous functions over time. Thus, (P1) involves an infinite number of optimization variables. Furthermore, $I(t)$'s are discrete optimization variables and the objective function $N$ is an integer-valued function. To summarize, (P1) is a non-convex optimization problem for which the optimal solution is difficult to obtain. Particularly, the new consideration of the number of handovers brings brand new challenges compared to existing works on the trajectory optimization for cellular-connected UAV (e.g., \cite{ShuowenCellular,ShuowenRadio,ShuowenOutage}). In the following, we will first transform the problem into a more tractable equivalent form by exploiting its structural properties.
\vspace{-4mm}
\section{Structural Properties of the Optimal Solution and Equivalent Problem Reformulation}\vspace{-1mm}
First, we introduce a set of auxiliary variables $\{T_i\}_{i=0}^{N}$, where $T_i, \ 1\leq i\leq N-1,$ denotes the time duration between the $i$-th handover and the $(i+1)$-th handover; $T_0$ represents the time duration from the mission start to the first handover; and $T_N$ denotes the time duration from the $N$-th handover to the mission completion. Moreover, we let an auxiliary vector $\bm{I}=[{I}_0,\cdots\!,{I}_i,\cdots\!,{I}_N]^T$ with ${I}_i\in \mathcal{M}, \ \forall i$ represent the \emph{GBS-UAV association sequence}, where ${I}_0=I(t),\ t\in[0,T_0]$, and ${I}_i=I(t),\ t\in\Big[\sum\limits_{j=0}^{i-1}{T_j}, \sum\limits_{j=0}^{i}{T_j}\Big], \ i=1,\cdots\!,N$. Based on this, (P1) can be equivalently transformed into \vspace{-3mm}
\begin{align}\mbox{(P2)}\ 
\mathop{\min_{\substack{T,\{\bm{u}(t),0\leq t\leq T\},\\ \bm{I}, \{T_i\}_{i=0}^N}}}\ &N\\[-1mm]
\mathrm{s.t.} \qquad \ \ &(\ref{P1-C1}), (\ref{P1-C4}), (\ref{P1-C5})\\[-1mm]
&\|\bm{u}(t)-\bm{g}_{I_{i}}\|-\bar d_{I_i}\le 0,\nonumber\\[-1mm]
&t\in \!\!\left[\sum_{j=0}^{i-1}T_j,\sum_{j=0}^{i}T_j\right]\!\!,\ i=0,\cdots\!,N\!\!\!\label{P2-C1}\\[-1mm]
&I_i\in \mathcal{M},\ i=0,\cdots\!,N.\label{P2-C2}
\end{align}
\vspace{-1mm}Note that compared to (P1), (P2) replaces the continuous-time $I(t)$ with a discrete sequence $\bm{I}$, based on which the number of handovers can also be interpreted as the cardinality of $\bm{I}$. However, there are still $M^{N+1}$ possible solutions of $\bm{I}$, which motivates us to provide the following proposition.

\begin{proposition}\label{Proposition_1}
The optimal number of handovers satisfies $N\leq M-1$. The optimal $\bm{I}$ to (P2) satisfies\vspace{-2mm}
\begin{align}
I_i\ne I_j, \ \forall i\ne j, \ i,j=0,\cdots\!,N.\label{proposition1}
\end{align}
\end{proposition}
\begin{IEEEproof}
Consider a feasible trajectory to (P2) denoted by $\{\hat{\bm{u}}(t),0\leq t\leq \hat{T}\}$ with repeated GBS-UAV associations, where the association sequence is denoted by $\hat{\bm{I}}=[\hat{I}_0,\cdots\!,\hat{I}_{k},\cdots\!,\hat{I}_{q},\cdots\!,\hat{I}_{\hat{N}}]^T$ with $k<q$ and $\hat{I}_{k}=\hat{I}_{q}$. We prove Proposition \ref{Proposition_1} by showing that a new feasible trajectory can be constructed without repeated association and with reduced number of handovers. Specifically, we construct the GBS-UAV association sequence by removing the $(k + 1)$-th to the $q$-th elements in $\hat{\bm{I}}$, which is given by $\tilde{\bm{I}}=[\hat{I}_0,\cdots\!,\hat{I}_{k},\hat{I}_{q+1},\cdots\!,\hat{I}_{\hat{N}}]^T$. The new feasible trajectory is then constructed by replacing the part in the original trajectory from $\hat{\bm{u}}(\sum_{j=0}^{k}\hat{T}_j)$ to $\hat{\bm{u}}(\sum_{j=0}^{q-1}\hat{T}_j)$ (during which the UAV leaves the association with GBS $\hat{I}_{k}$ and becomes associated with it again) by a straight-line path with maximum speed from $\hat{\bm{u}}(\sum_{j=0}^{k}\hat{T}_j)$ to $\hat{\bm{u}}(\sum_{j=0}^{q-1}\hat{T}_j)$. Note that since $\hat{\bm{u}}(\sum_{j=0}^{k}\hat{T}_j)$ and $\hat{\bm{u}}(\sum_{j=0}^{q-1}\hat{T}_j)$ are both within the disk-shaped coverage region of GBS $\hat{I}_{k}$, the newly constructed trajectory satisfies the communication quality constraint. Moreover, the replaced part consumes no longer time compared to the original part, as it incurs shortest distance with maximum speed, thus the newly constructed trajectory satisfies the mission completion time constraint. Finally, the newly constructed trajectory yields a smaller number of handovers of $\tilde{N}=\hat{N}-q+k< \hat{N}$. This thus completes the proof of Proposition \ref{Proposition_1}.
\end{IEEEproof}

Proposition \ref{Proposition_1} indicates there should be no more than $M-1$ handovers in the UAV's mission, and significantly reduces the feasible set of $\bm{I}$ based on (\ref{proposition1}). Furthermore, we let $\bm{u}_i=\bm{u}\Big(\sum\limits_{j=0}^{i-1}{T_{j}}\Big)$ represent the location where the UAV is handed over from GBS $\bm{g}_{I_{i-1}}$ to GBS $\bm{g}_{I_{i}}$. For completeness, we define the initial location $\bm{u}_0=\bm{u}(0)=\bar{\bm{u}}_0$ and the final location $\bm{u}_{N+1}=\bm{u}(T)=\bar{\bm{u}}_{\mathrm{F}}$. Based on $\{\bm{u}_i\}_{i=0}^{N+1}$ defined above, we propose the following proposition.

\begin{proposition}\label{Proposition_2}
Without loss of optimality, the optimal solution to (P2) can be assumed to satisfy the following conditions:\vspace{-2mm}
\begin{align}
&T_i=\frac{\|\bm{u}_{i+1}-\bm{u}_{i}\|}{V_{\max}},\ i=0,\cdots\!, N\label{proposition2-1}\\[-1mm]
&\bm{u}(t)=\bm{u}_{i}+\Big(t-\sum\limits_{j=0}^{i-1}{T_{j}}\Big)V_{\max}\frac{\bm{u}_{i+1}-\bm{u}_{i}}{\|\bm{u}_{i+1}-\bm{u}_{i}\|},\nonumber\\[-1mm]
&\qquad t\in \left[\sum\limits_{j=0}^{i-1}{T_{j}}, \sum\limits_{j=0}^{i}{T_{j}}\right],\ i=0, \cdots\!, N\label{proposition2-2}.
\end{align}
\end{proposition}
\begin{IEEEproof}
We prove Proposition \ref{Proposition_2} by showing that for any feasible solution to (P2) denoted by $(\hat T, \{\hat{\bm{u}}(t),0\leq t\leq \hat{T}\}, \bm{I},\{\hat{T}_i\}_{i=0}^{N})$ that does not satisfy the conditions in (\ref{proposition2-1}), (\ref{proposition2-2}), we can always construct a feasible solution to (P2) denoted by $(T, \{\bm{u}(t),0\leq t\leq T\}, \bm{I},\{T_i\}_{i=0}^N)$ that satisfies these conditions with the same objective value. We set the handover locations in $\{{\bm{u}(t)}, 0\leq t \leq T\}$ as the same ones in $\{\hat{\bm{u}}(t), 0\leq t \leq \hat{T}\}$, i.e., $\bm{u}_i=\hat{\bm{u}}\Big(\sum\limits_{j=0}^{i-1}{\hat{T}_{j}}\Big), \ i=0, \cdots\!, N+1$. Based on this, the new solution is constructed based on (\ref{proposition2-1}), (\ref{proposition2-2}). Note that $\hat{T}_{i}$ denotes the time duration for the UAV to fly from $\bm{u}_{i+1}$ to $\bm{u}_{i}$, where $\hat{T}_{i}=\frac{\| \bm{u}_{i+1}-\bm{u}_{i}\|}{\|\dot{\bm{u}}(t)\|}$, thus $\hat{T}_{i}\geq \frac{\| \bm{u}_{i+1}-\bm{u}_{i}\|}{V_{\max}}$ should hold due to the constraint in (\ref{P1-C4}). By noting that ${T}_{i}= \frac{\| \bm{u}_{i+1}-\bm{u}_{i}\|}{V_{\max}}$, we have ${T}_{i}\leq \hat{T}_{i},\ i=0, \cdots\!, N$, and consequently ${T}=\sum\limits_{i=0}^{N}{\frac{\| \bm{u}_{i+1}-\bm{u}_{i}\|}{V_{\max}}}\leq \hat{T}=\sum\limits_{i=0}^{N}{\frac{\| \bm{u}_{i+1}-\bm{u}_{i}\|}{\|\dot{\bm{u}}(t)\|}}\leq T_{\max}$. Moreover, the newly constructed solution also satisfies the communication quality constraint since the line segment between two consecutive handover locations is guaranteed to lie in the disk-shaped GBS coverage region. Thus, the new solution is feasible for (P2) with unchanged objective value $N$, which completes the proof of Proposition \ref{Proposition_2}.
\end{IEEEproof}

Proposition \ref{Proposition_2} indicates that the optimal horizontal trajectory is composed of multiple connected line segments, and the UAV flies along these line segments at maximum speed $V_{\max}$. Specifically, the $i$-th line segment's start and end points are the $(i-1)$-th handover location $\bm{u}_{i-1}$ and the $i$-th handover location $\bm{u}_{i}$, respectively. Based on this, we equivalently transform (P2) and consequently (P1) into the following problem:\vspace{-3mm}
\begin{align}\mbox{(P3)}\ 
\mathop{\min_{\{\bm{u}_i\}_{i=0}^{N+1},\bm{I}}} \ &N\\[-1.5mm]
\mathrm{s.t.} \quad \ &(\ref{P1-C1}),(\ref{P2-C2}),(\ref{proposition1})\\[-1mm]
&\bm{u}_0 =\bar{\bm{u}}_0,\ \bm{u}_{N+1} =\bar{\bm{u}}_{\mathrm{F}}\label{P3-C1}\\[-1mm]
&\sum\limits_{i=0}^{N}{\| \bm{u}_{i+1}-\bm{u}_{i}\|}\leq T_{\max}V_{\max}\label{P3-C2}\\[-1mm]
&\| \bm{u}_i-\bm{g}_{I_{i}}\|\leq \bar d_{I_{i}},\ i=0,\cdots\!, N\label{P3-C3}\\[-1mm]
&\| \bm{u}_i-\bm{g}_{I_{i-1}}\| = \bar d_{I_{i-1}},\ i=1,\cdots\!, N\label{P3-C4}\\[-1mm]
&\| \bm{u}_{N+1}-\bm{g}_{I_{N}}\|\leq\bar d_{I_{N}}.\label{P3-C5}
\end{align}

Note that in contrast to (P1), the equivalent problem (P3) is a joint optimization problem of the GBS-UAV association sequence and handover locations, which does not involve any continuous function over time. However, it is still a non-convex optimization problem due to the non-convex constraint in (\ref{P3-C4}) and the discrete optimization variables in $\bm{I}$. In the following, we will leverage \emph{graph theory} to handle (P3).
\vspace{-4mm}
\section{Proposed Solution to Problem (P3)}\vspace{-1mm}
Note that a key difficulty in (P3) lies in the mixture of the continuous and discrete optimization variables representing the handover locations and GBS-UAV association sequence, respectively. To address this challenge, we first propose an effective structural design of the handover locations, based on which (P3) can be modeled and tackled via graph theory.\vspace{-4mm}
\subsection{Handover Location Design}\vspace{-1mm}
Motivated by Proposition \ref{Proposition_2} and the fact that the communication quality improves as the UAV flies closer to the GBS, we propose a structural handover location design. Specifically, the UAV firstly flies from the start location to the location on top of the firstly connected GBS. Then, the UAV flies in a straight line between the locations on top of the sequentially associated GBSs, where the $i$-th (horizontal) handover location $\bm{u}_i$ is located at the intersection of the line segment from the $I_{i-1}$-th to the $I_i$-th GBSs and the coverage boundary of the $I_{i-1}$-th GBS. Finally, the UAV flies from the location above the lastly associated GBS to the final destination, as illustrated in Fig. \ref{fig_1}. We let $\tilde{T}_{i}, \ 1\leq i\leq N,$ represent the time duration for the UAV to fly between the locations on top of GBSs $I_{i-1}$ and $I_{i}$; $\tilde{T}_{0}$ denote the time duration for the UAV to fly from the initial location to the location above the firstly associated GBS; $\tilde{T}_{N+1}$ denote the time duration for the UAV to fly from the location above the lastly associated GBS to the final destination. The UAV's trajectory and mission completion time are thus given by\vspace{-3mm}
\begin{align}
\!\bm{u}(t)=\left\{
\begin{array}{l}
\bm{u}_0+tV_{\max}\frac{\bm{g}_{I_0}-\bm{u}_{0}}{\|\bm{g}_{I_0}-\bm{u}_{0}\|},\ t\in [0,\tilde{T}_{0}]\\
\bm{g}_{I_0}+\Big(t-\sum\limits_{j=0}^{i}{\tilde{T}_{j}}\Big)V_{\max}\frac{\bm{g}_{I_{i+1}}-\bm{g}_{I_i}}{\|\bm{g}_{I_{i+1}}-\bm{g}_{I_i}\|},\\ \quad t\in \left[\sum\limits_{j=0}^{i}{\tilde{T}_{j}},\sum\limits_{j=0}^{i+1}{\tilde{T}_{j}} \right],\ i=0,\cdots\!,N\!-\!1\\
\bm{g}_{I_N}+\Big(t-\sum\limits_{j=0}^N{\tilde{T}_{j}}\Big)V_{\max}\frac{\bm{u}_{N+1}-\bm{g}_{I_N}}{\| \bm{u}_{N+1}-\bm{g}_{I_N}\|},\\ \quad t\in \left[\sum\limits_{j=0}^N{\tilde{T}_{j}},T\right]
\end{array} \right.,\label{Handoverdesign-1}
\end{align}
\vspace{-2mm}
\begin{align}
\!T=\frac{\| \bm{u}_0 -\bm{g}_{I_0}\|}{V_{\max}}\!+\!\frac{\| \bm{u}_{N+1}\!-\!\bm{g}_{I_N} \|}{V_{\max}}\!+\!\sum_{i=0}^{N-1}{\frac{\| \bm{g}_{I_{i+1}}\!-\!\bm{g}_{I_{i}}\|}{V_{\max}}}.\label{Handoverdesign-2}
\end{align}

Based on the above handover location structure, (P3) can be transformed to the following problem:\vspace{-2mm}
\begin{align}\!\!\mbox{(P4)}\ 
\mathop{\min_{\bm{I}}} \ & N\\[-2mm]
\mathrm{s.t.} \ &(\ref{P1-C1}), (\ref{P2-C2}), (\ref{proposition1})\\[-1mm]
&\|\bm{u}_0\!-\!\bm{g}_{I_0}\|\!+\!\|\bm{u}_{N+1}\!-\!\bm{g}_{I_N}\|\!+\!\sum_{i=0}^{N-1}{\|\bm{g}_{I_{i+1}}\!-\!\bm{g}_{I_{i}}\|}\nonumber\\[-1mm]
&\qquad\qquad\qquad\qquad\qquad\qquad\leq T_{\max}V_{\max}\label{P4-C1}\\[-1mm]
&\| \bm{g}_{I_{i}}\!-\!\bm{g}_{I_{i-1}}\|\!-\!\bar d_{I_i}\!-\!\bar d_{I_{i-1}}\le 0,\ i\!=\!1,\cdots\!,N.\label{P4-C2}
\end{align}
Notice that the only variable in (P4) is the GBS-UAV association sequence $\bm{I}$, which should satisfy the constraint in (\ref{P4-C2}) such that each pair of GBSs $I_{i-1}$ and $I_{i}$ can be consecutively associated with the UAV. Although the optimal solution to (P4) may not be optimal to (P3) and consequently (P1) due to the specific handover location design considered, it enables an equivalent graph-based modeling, as shown below.
\vspace{-5mm}
\subsection{Equivalent Graph-Based Model and Solution for (P4)}\vspace{-1mm}

First, we present an equivalent graph-based model of (P4). We construct a graph $G=(V,E)$, where the vertex set is given by $V=\{ U_0,G_{1},\cdots\!,G_{M},U_{\mathrm{F}}\}$, each representing a GBS or the UAV's initial/final location. The edge set is given by\vspace{-2mm}
\begin{align}
E&=\{(U_0,G_m ):\|\bar{\bm{u}}_0-\bm{g}_m\|\le\bar d_m,\ m\in\mathcal{M}\} \nonumber
\\[-1mm]
&\cup \{(G_m,G_n):\|\bm{g}_m-\bm{g}_n\|\!\le\!\bar d_{m}\!+\!\bar d_{n},\ m,n\in\mathcal{M}, \ m\!\ne\!n\} \nonumber
\\[-1mm]
&\cup \{(U_{\mathrm{F}},G_m):\|\bar{\bm{u}}_{\mathrm{F}}-\bm{g}_m\| \le \bar d_m,\ m\in\mathcal{M}\}.\label{E}
\end{align}
Note that an edge exists between two vertices if and only if their corresponding GBSs can be potentially consecutively associated with the UAV, or the UAV can be associated with a GBS at the start or the end of the mission. We define a novel \emph{handover weight} of each edge, which is given by\vspace{-2mm}\begin{align}
&W_{\mathrm{H}}(U_0,G_m)= 0, W_{\mathrm{H}}(G_m,G_n)= 1,\nonumber
\\[-1mm]
&W_{\mathrm{H}}(U_{\mathrm{F}},G_m)= 0,\ m, n\in \mathcal{M},\ m\ne n.\label{WH}
\end{align}
Moreover, we define a \emph{distance weight} of each edge given by\vspace{-2mm}
\begin{align}
&W_{\mathrm{D}}(U_0,G_m)=\|\bar{\bm{u}}_0-\bm{g}_m\|,
W_{\mathrm{D}}(G_m,G_n)=\|\bm{g}_m-\bm{g}_n\|,\nonumber
\\[-1mm]
&W_{\mathrm{D}}(U_{\mathrm{F}},G_m) =\|\bar{\bm{u}}_{\mathrm{F}}-\bm{g}_m\|,
\ m, n\in \mathcal{M},\ m\ne n.\label{WD}
\end{align} Note that any path from $U_0$ to $U_{\mathrm{F}}$ denoted by $(U_0,G_{I_0},G_{I_1},\cdots\!,G_{I_N},U_{\mathrm{F}})$ represents a feasible GBS-UAV association sequence $\bm{I}=[I_0,\cdots\!,I_N]^T$, based on which the communication quality constraint can be satisfied with the proposed handover location design. The corresponding total number of handovers can be represented as $N=f_{\mathrm{H}}(\bm{I})\!=\!{W_{\mathrm{H}}(U_0,I_0)}+\sum\limits_{i=0}^{N-1}{W_{\mathrm{H}}(I_{i+1},I_i)}+{W_{\mathrm{H}}(I_N,U_{\mathrm{F}})}$, and the corresponding mission completion time can be represented as $f_{\mathrm{T}}(\bm{I})=({W_{\mathrm{D}}(U_0,I_0)}+\sum\limits_{i=0}^{N-1}{W_{\mathrm{D}}(I_{i+1},I_i)}+{W_{\mathrm{D}}(I_N,U_{\mathrm{F}})})/V_{\max}$. Thus, (P4) can be equivalently expressed as\vspace{-5mm}
\begin{align}\mbox{(P5)}\ 
\min_{\bm{I}\in \kappa}\ &f_{\mathrm{H}}(\bm{I})\\[-1mm]
\mathrm{s.t.} \ &f_{\mathrm{T}}(\bm{I})\leq T_{\max},\label{P5-C1}
\end{align}
where $\kappa=\{[{I}_0,\cdots\!,{I}_N]^T:{I}_i\in \mathcal{M},\ I_i\ne I_j,\ \forall i\ne j\}.$ Note that (P5) is still a non-convex optimization problem belonging to the class of \emph{constrained shortest path problems} (or weight constrained shortest path problems), which has been shown to be NP-hard \cite{NP}. To tackle this problem, we propose an efficient algorithm based on Lagrangian relaxation and graph theory \cite{ShuowenOutage}. Specifically, the Lagrangian of (P5) is given by $\mathscr{L}(\bm{I},\lambda)=f_{\mathrm{H}}(\bm{I})+\lambda (f_{\mathrm{T}}(\bm{I})-T_{\max}),$ where $\lambda\geq 0$ denotes the dual variable associated with the constraint in (\ref{P5-C1}). The Lagrange dual function is then given by $g(\lambda)=\min\limits_{\bm{I}}{\mathscr{L}(\bm{I}, \lambda)}=\min\limits_{\bm{I}} f_{\mathrm{H}}(\bm{I})+\lambda(f_{\mathrm{T}}(\bm{I})-T_{\max})$. Consequently, the dual problem of (P5) is given by\vspace{-2mm}
\begin{align}\mbox{(P5-Dual)}\ 
\max_{\lambda\geq0}\ \min_{\bm{I}\in \kappa} \ f_{\mathrm{H}}(\bm{I})+\lambda (f_{\mathrm{T}}(\bm{I})-T_{\max}).
\end{align} 
(P5-Dual) is a convex optimization problem which can be solved by iteratively updating $\lambda$ via the subgradient method. Due to the non-convexity of (P5), the duality gap between (P5) and (P5-Dual) is generally non-zero. To reduce this gap, we apply the $K$-shortest path method to obtain $K$ best solutions for minimizing $f_{\mathrm{H}}(\bm{I})+\lambda^\star f_{\mathrm{T}}(\bm{I})$ via Yen's algorithm, where $\lambda^\star$ is the optimal solution to (P5-Dual). Then, the final solution of $\bm{I}$ is selected as the best solution among the optimal $\bm{I}$ to (P5-Dual) and the $K$ additional solutions. More details  can be found from Appendix E of \cite{ShuowenOutage} and are omitted due to limited space. The worst-case complexity of this algorithm is $\mathcal{O}(M^4\log^2M+M^3K)$, which is much lower than that for solving (P5) via exhaustive search, i.e., $O(M!)$ \cite{ShuowenOutage}.\vspace{-4mm}
\section{Numerical Results}
\begin{figure}[t]
\centering
\includegraphics[width=0.30\textwidth]{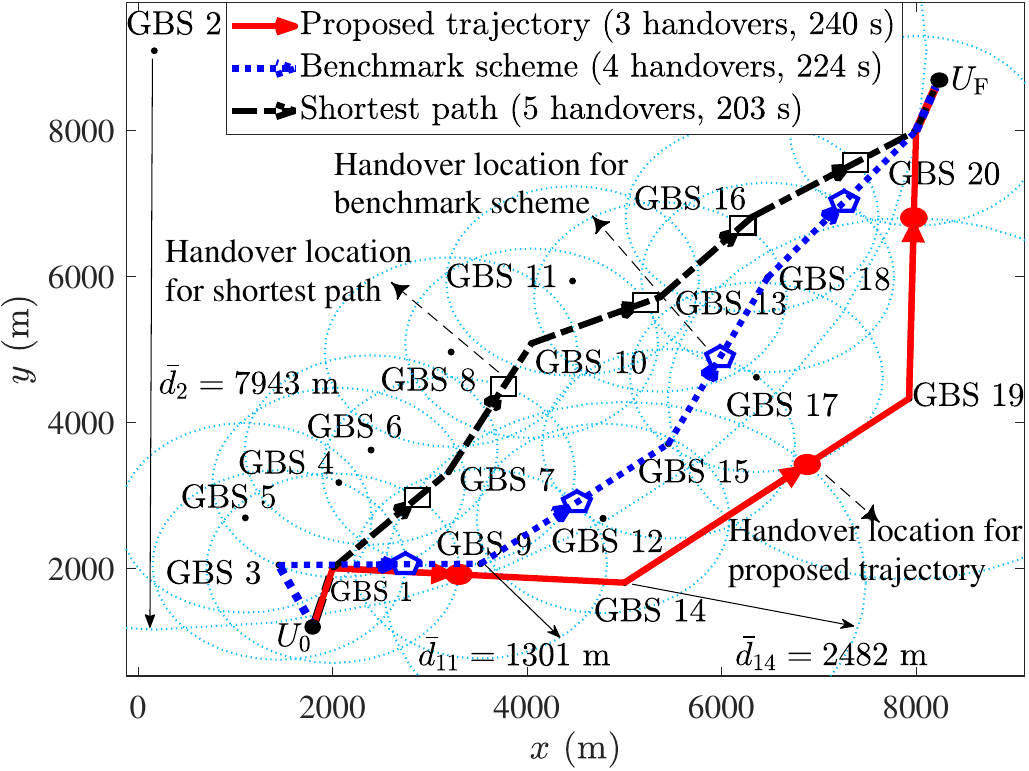}\vspace{-2.6mm}
\caption{Illustration of different trajectory designs.}\vspace{-8mm}
\label{fig_2}
\end{figure}

\begin{figure}[t]
	\centering
	\begin{subfigure}[b]{0.45\linewidth}
		\includegraphics[width=\linewidth]{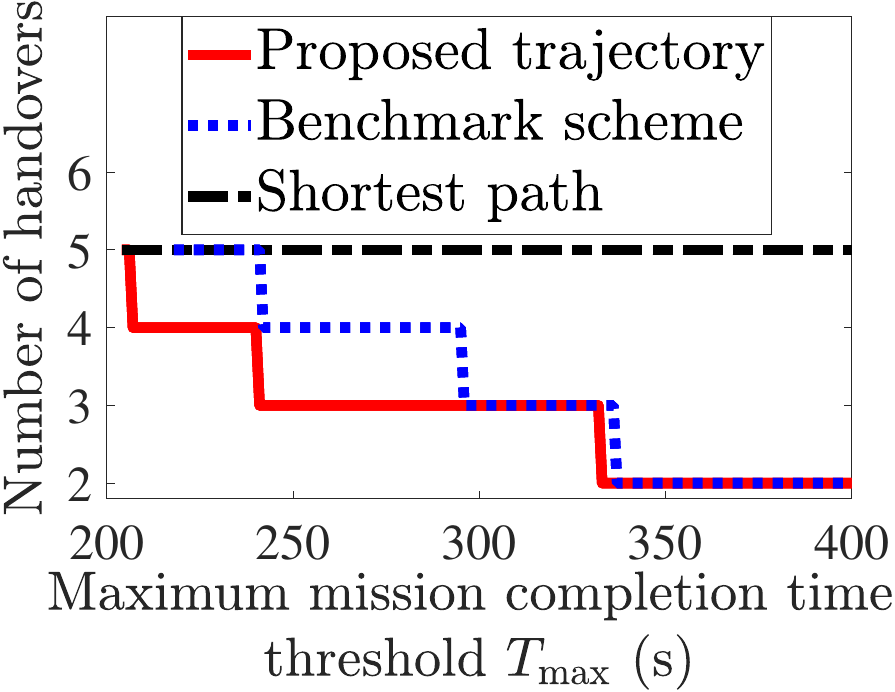}\vspace{-2mm}
		\caption{No. of handovers vs. $T_{\max}$.}\vspace{-2mm}
		\label{fig3:sub1}
	\end{subfigure}
	\hfill
	\begin{subfigure}[b]{0.45\linewidth}
		\includegraphics[width=\linewidth]{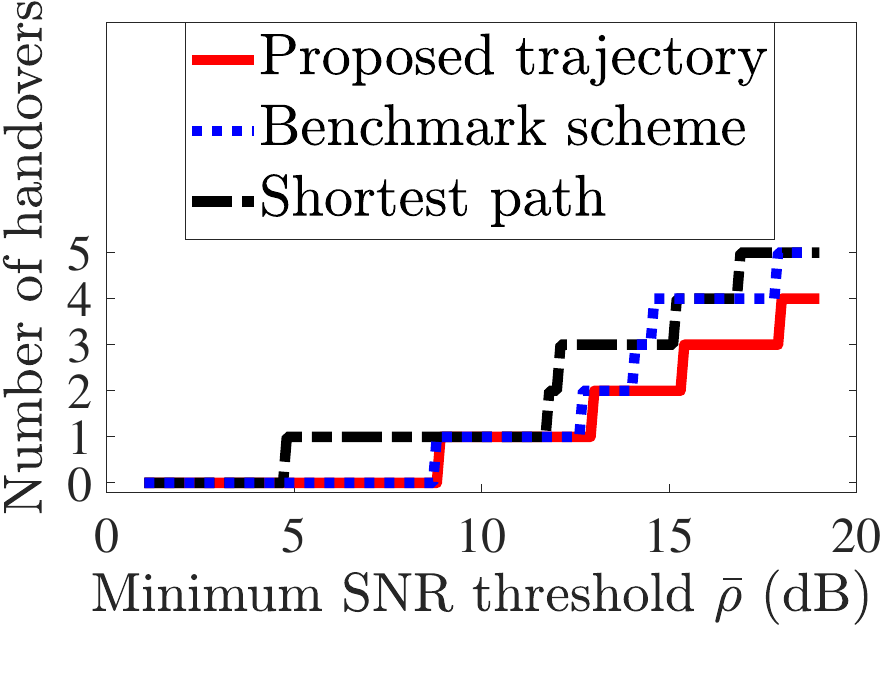}\vspace{-2mm}
		\caption{No. of handovers vs. $\bar{\rho}$.}\vspace{-2mm}
		\label{fig3:sub2}
	\end{subfigure}
	\caption{Illustration of the trade-off between the number of handovers and mission/communication requirements.}\vspace{-7mm}
	\label{fig:combined}
\end{figure}

\vspace{-2mm}
In this section, we provide numerical results to evaluate the performance of our proposed handover-aware trajectory design. We set $V_{\max}=50$ m/s, $\sigma^2=-90$ dBm, $\beta_0=-30$ dB, and $H=90$ m. In Fig. \ref{fig_2}, we randomly generate the locations of $M=20$ GBSs in a $10\times 10\ \mathrm{km}^2$ square region. We consider one large GBS indexed by $2$ with transmit power $35.7$ dBm and antenna height $20$ m, two medium GBSs indexed by $14$ and $19$ with transmit power $25.6$ dBm and antenna height $15$ m, and $17$ small GBSs with transmit power $20$ dBm and antenna height $12.5$ m \cite{3GPP}.
For comparison, we consider a shortest path trajectory which aims to minimize the mission completion time subject to a receive SNR constraint $\bar{\rho}$ using Method I in \cite{ShuowenCellular} based on a similar handover location design as in Section V-A. Moreover, we consider a benchmark scheme which aims to solve (P5) via the genetic algorithm \cite{Genetic}. Under a setup of $T_{\max}\!=\!270$ s and $\bar{\rho}\!=\!17.7$ dB, we show the trajectories via the proposed handover-aware design, the shortest path trajectory, and the benchmark scheme in Fig. \ref{fig_2}. It is observed that the proposed design yields a GBS-UAV association sequence of $[1,\!14,\!19,\!20]^T$, while the shortest path trajectory and the benchmark scheme yield sequences with more handovers, i.e., $[1,7,10,13,16,20]^T$ and $[1,9,15,18,20]^T$, respectively. Furthermore, under $\bar{\rho}\!=\!17.7$ dB, we show in Fig. \ref{fig:combined} (a) the number of handovers versus the mission completion time threshold $T_{\max}$ for different trajectories. It is observed that the handover performance improves as increased time is allowed for the UAV's flight, which enables higher design flexibility. In contrast, the shortest path  trajectory yields a constant number of handovers which is much larger compared to the proposed design, and the benchmark scheme is also outperformed by the proposed design. Under $T_{\max}\!=\!270$ s, we show in Fig. \ref{fig:combined} (b) the number of handovers versus the minimum SNR threshold $\bar{\rho}$. It is observed that all three designs yield larger numbers of handovers as the communication quality constraint becomes more stringent, while our proposed design still outperforms the other two schemes, thus validating its effectiveness.

\vspace{-4mm}
\section{Conclusions}\vspace{-1mm}
Considering a cellular-connected UAV with a mission of flying between two locations, this paper studied the trajectory optimization to minimize the number of GBS handovers along the flight, subject to a communication quality constraint and a maximum mission completion time constraint. The formulated optimization problem was non-convex and involved infinite optimization variables. By exploiting the problem structure and leveraging graph theory as well as Lagrange duality, a polynomial-time algorithm was proposed to find an approximate solution. Numerical results showed that the proposed trajectory design effectively reduces the number of handovers compared with various benchmark schemes.
\vspace{-4mm}

\bibliographystyle{IEEEtran}
\bibliography{reference}

\end{document}